# *Ab-initio* Study on the Magnetic Structures in the Ordered Mn$_3$Pt Alloy

Yohei Kota, Hiroki Tsuchiura and Akimasa Sakuma

Department of Applied Physics, Graduate School of Engineering, Tohoku University,
Aoba 6-6-05, Aoba-ku, Sendai, Miyagi 980-8579, Japan

**We study the electronic states of the magnetically ordered Mn$_3$Pt alloy within the density functional theory. Mn$_3$Pt has been believed that one third of Mn atoms have no magnetic moment in an antiferromagnetic phase (so-called the F-phase) realized in the temperature range of 400 K < *T* < 475 K. We show that this experimentally suggested spin configuration is energetically so much unfavorable that it would be irrelevant to the F-phase. We discuss the possibility that the spin moments on the one third of Mn atoms are not paramagnetic but thermally fluctuating in the F-phase. The present results have an immediate connection with the recent neutron scattering study [T. Ikeda and Y. Tsunoda, J. Phys. Soc. Jpn., vol. 72, pp. 2614—2621, October. 2003.].**

## I. INTRODUCTION

GEOMETRICALLY frustrated antiferromagnetic (AF) systems have been extensively studied because they show many unusual behaviors in magnetic and thermal properties. Their classical ground state has a thermodynamic degeneracy with a potential instability to various magnetic phases. In general, thermal fluctuations may drive symmetry breaking to the phase with the largest entropy and lead to the possibility of phase transitions between different magnetic phases.

The magnetically ordered Mn$_3$Pt alloy with the L1$_2$ (Cu$_3$Au) type chemical structure has long been studied and considered to be such a frustrated AF system, which shows the first order phase transition between two AF phases [1], [2]. The magnetic structures of these two phases are determined by neutron scattering measurements. In the low temperature phase, so called the D-phase realized below $T_t$ = 400 K, the spin moments of the three Mn atoms in the unit cell order in a triangular arrangement in the (111) plane, each spin pointing to the center of the triangle, shown in Fig. 1 (a). In the high temperature phase, so-called the F-phase realized in the temperature range of $T_t < T < T_N$ = 475 K, in contrast, a collinear magnetic structure shown in Fig. 1 (b) is proposed. Notice that the magnetic unit cell is double atomic cell in this magnetic structure. A peculiar feature of the F-phase is that one third of Mn atoms on the sites where the magnetic mean-field is cancelled have no magnetic moments, as shown in Fig. 1 (b). We will call these sites as "B"-sites, and other Mn sites as "S"-sites.

Theoretically, J. Kübler, K.-H. Höck, J. Sticht, and A. R. Williams studied some possible magnetic structures in Mn$_3$Pt within the augmented spherical wave (ASW) method, in their pioneering work for density functional theory (DFT) of noncollinear magnetism [3]. The D-phase magnetic structure (Fig. 1 (a)) was confirmed to be the ground state for Mn$_3$Pt with a magnetic moment of 2.93 μ$_B$ per atom. By contrast, no electronic structure calculations for the collinear structure proposed for the F-phase has been carried out and little is known theoretically about the spin configuration of Mn atoms on the B sites.

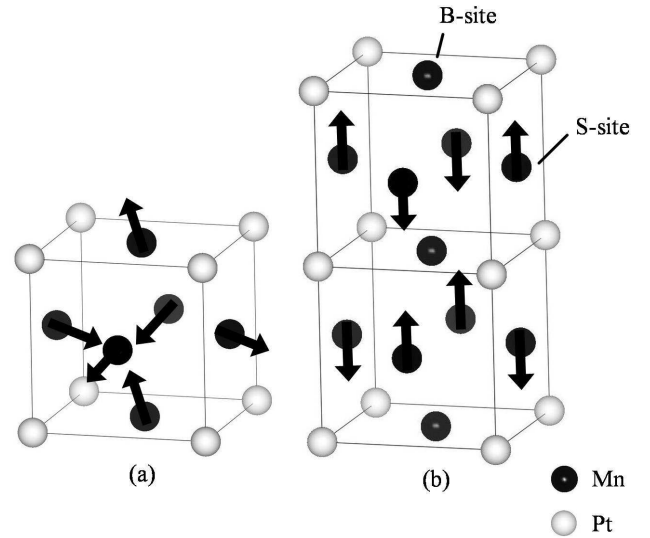

Fig. 1. Experimentally proposed magnetic structures for the ordered Mn$_3$Pt alloy in (a) the D-phase and (b) the F-phase [1]. These figures are plotted by using VESTA [5].

Recently, it has been reported that inelastic diffuse scattering located along the first Brillouin zone boundary of the fcc-structure is observed in neutron scattering measurements in the F-phase, implying the Mn spins on the B-sites are not paramagnetic but strongly fluctuating [4]. To understand the nature of the phase transition between the D-phase and F-phases, it is important to reveal whether the Mn atoms on the B-sites in the F-phase have magnetic moments or not. Motivated by these backgrounds, we study the electronic states of various magnetic structures in the ordered Mn$_3$Pt alloy by using *ab-initio* calculation techniques based on the DFT.

We begin below by introducing the computational details of the calculation using DFT. We then present our results on various magnetic structures for Mn$_3$Pt. We show that the collinear structure proposed for the F-phase is energetically highly unfavorable. Finally we discuss the possibility that the Mn spins on the B-sites is thermally fluctuated in the temperature range of the F-phase.

## II. COMPUTATIONAL DETAILS

In this work, DFT calculations have been used to determine all energetic and electronic results. The Kohn-Sham equations are solved self-consistently in a plane-wave basis set, in conjunction with the generalized gradient approximation (GGA) projector augmented wave (PAW) potentials, as implemented in the Vienna *ab-initio* simulation package (VASP) [6], [7]. We use the exchange-correlation functional determined by Ceperly and Alder and parameterized by Perdew and Zunger.

In all the calculations using VASP, the kinetic energy cutoff is 270 eV. The k-point mesh used for the bulk calculation is $(11 \times 11 \times 11)$ in the first Brillouin zone. The total energy convergence is less than 0.1meV per unit cell. The lattice constants are set to experimental values of a = b = c = 3.833 Å for the D-phase [1], while a = b = 3.882 Å and c = 3.876 Å for the F-phase [4].

We also employ the tight-binding (TB) linear muffin-tin orbital (LMTO) method in the local density approximation (LDA) combined with the coherent potential approximation (CPA) to estimate the fluctuation effects of the Mn spins at the B-sites on the electronic energy [8], [9].

## III. RESULTS AND DISCUSSION

First, let us look at the electronic ground state of the Mn$_3$Pt. We carry out calculations for three different magnetic structures shown in Fig. 2 (I) - (III), all of which are consistent with the experimental results obtained in the D-phase, that is, the moments of the three Mn atoms in the unit cell order in a triangular arrangement in the (111) plane [1]. Taking account of the spin-orbit interaction to determine the angle between the Mn spins and crystal axis, we actually obtain the type (I) in Fig. 2 magnetic structure as the lowest energy state of the system as shown in Table I. The magnetic moment obtained here is 3.09 $\mu_B$ per Mn atom, which is in

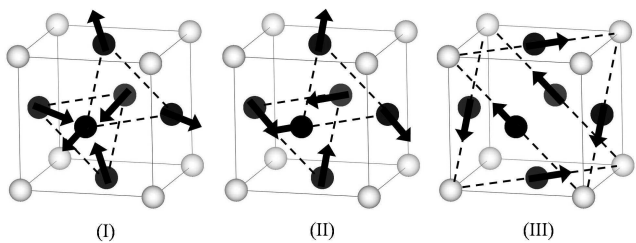

Fig. 2. Possible magnetic structures consistent with the experimental results in the D-phase.

TABLE I
ENERGIES IN THE D-PHASE

| Structure | E (eV) | ΔE (meV) |
|---|---|---|
| I | -34.2310 | 0 |
| II | -34.2295 | +1.5 |
| III | -34.2282 | +2.8 |

Calculated electronic energy per unit cell for the magnetic structures shown in Fig. 2.

fairly good agreement with the experimental value $3.0 \pm 0.3$ $\mu_B$ [2]. We note here that the present results are consistent with the previous theoretical work done by J. Kübler, K.-H. Höck, J. Sticht, and A. R. Williams [3].

Next we study the magnetic structures with higher energies, which may be relevant to the F-phase. There is an experimental consensus on the magnetic structure in the F-phase, that is, the magnetic moments on the Mn atoms on the

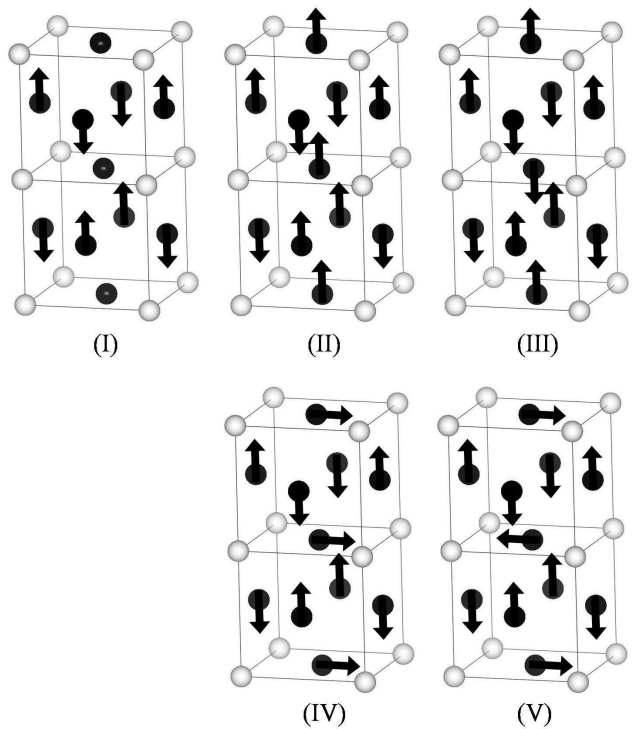

Fig. 3. Various magnetic structures as candidates for the F-phase. The magnetic moments of the Mn atoms at the S-sites order in a collinear AF arrangement.

TABLE II
ENERGIES IN THE F-PHASE

| Structure | E (eV) | ΔE (meV) |
|---|---|---|
| D-phase | -33.685 | 0 |
| I | -32.582 | +1103 |
| II | -33.529 | +158 |
| III | -33.440 | +245 |
| IV | -33.464 | +221 |
| V | -33.444 | +241 |

Calculated electronic energy per unit cell (Mn$_3$Pt) for the magnetic structures shown in Fig. 3.

S-sites order in a collinear AF arrangement as illustrated in Fig. 1 (b). Relying on this fact, here we study the various magnetic configurations with collinear AF structure for the S-sites. Five representative configurations are shown in Fig. 3 (I) - (V). The calculated energies are summarized in Table II. First, we point out that the configuration (I) in Fig. 3, which is originally proposed for the F-phase, has much higher energy than not only the ground state but also all the other states in Fig. 3. Thus we may conclude that the configuration (I) in Fig. 3 is irrelevant to the F-phase realized in the temperature range of 400 K $< T <$ 475 K. In other words, the Mn spins on the B-sites would have magnetic moments even in the F-phase. This result strongly supports the recent experimental results obtained by T. Ikeda and Y. Tsunoda [4].

Here we must consider the next question of the magnetic structure realized in the F-phase. As shown in Table II, the configuration (II) in Fig. 3 has the lowest energy among the magnetic structures having the collinear S-site moments. This structure, contrary to the experimental observation for the F-phase, has a net magnetic moment and a resulting magnetization. Thus, the question to address is whether the Mn spins on the B-sites in the configuration (II) in Fig. 3 can thermally fluctuate in the temperature region of the F-phase.

As a first step toward this issue, we study the effects of the disordered spin configuration of the Mn spins on the B-sites on the electronic energy of the system, as a simple mimic of the thermally fluctuating spins. To treat such a disordered local moment state, we employ the TB-LMTO method combined with the CPA [8] - [10]. The disorder is treated by replacing the magnetic unit cell as usual the average scattered determined self-consistently. We show the three disorder patterns involved in the present calculation in Fig. 4 (I) - (III), and the calculated energy in Table III. Note here that the Mn spins on the S-sites are still collinearly ordered. The energy losses due to the random configuration of the Mn spins on the B-sites are estimated in the order of meV per unit cell (Mn$_3$Pt). We should note that the TB-LMTO method with the CPA may not give reliable numbers to the nearest meV. However, we believe that the present result implies that the Mn spins on the B-sites are able to fluctuate thermally in the temperature region of the F-phase with $k_B T \sim$ 30 meV per atom, with the large entropy gain.

Last, we note that, it is useful to evaluate the spin-stiffness or the exchange coupling constants for the Mn spins based on the *ab-initio* calculations to reveal the thermal fluctuation effects or to estimate the $T_N$ and $T_t$. More detailed studies including these results are being carried out.

## IV. SUMMARY

We have investigated the magnetic structures of the D-phase and F-phases of the ordered Mn$_3$Pt alloy within the DFT. Our analysis has shown that the one third of Mn atoms in the F-phase is not paramagnetic and has suggested that they are thermally fluctuating, being consistent with the recent experimental data [4].

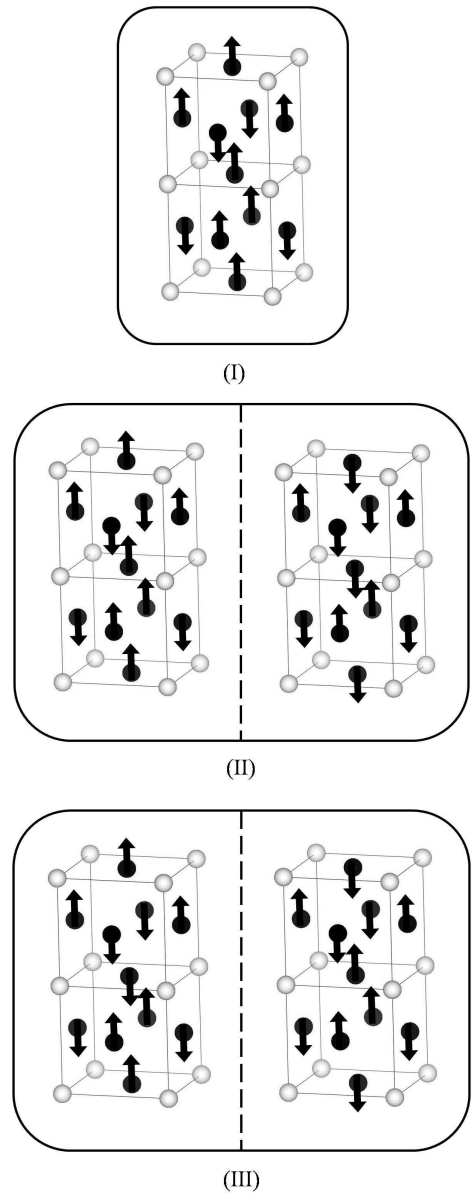

Fig. 4. The patterns of disorder studied within the CPA.

TABLE III
ENERGIES GIVEN BY CPA

| Pattern | E (eV) | ΔE (meV) |
|---|---|---|
| I | -2015.1455 | 0 |
| II | -2015.1444 | +1.1 |
| III | -2015.1442 | +1.3 |

Calculated electronic energy per unit cell (Mn$_3$Pt) for the disordered phases shown in Fig. 4.

ACKNOWLEDGMENT

Numerical computation in this work was partially carried out by using supercomputing resources at Information Synergy Center, Tohoku University and the Yukawa Institute Computer Facility. This work is supported by Grant-in-Aids